\documentclass[11pt]{article}
\pdfoutput=1
\usepackage{amsmath,amssymb,color,epsfig,cite}
\usepackage{graphicx}
\usepackage{subfigure}
\usepackage{setspace}
\usepackage{amsthm,mathenv}
\usepackage{ulem}

\allowdisplaybreaks[4]

\usepackage{amsfonts}

\newcommand{\hoch}[1]{$\, ^{#1}$}

\makeatletter
\@addtoreset{equation}{section}
\makeatother

\newcommand{\be}{\begin{equation}}
\newcommand{\ee}{\end{equation}}
\newcommand{\bea}{\setlength\arraycolsep{2pt} \begin{eqnarray}}
\newcommand{\eea}{\end{eqnarray}}

\def\0{{\sst{(0)}}}
\def\1{{\sst{(1)}}}
\def\2{{\sst{(2)}}}
\def\3{{\sst{(3)}}}
\def\4{{\sst{(4)}}}
\def\5{{\sst{(5)}}}
\def\6{{\sst{(6)}}}
\def\7{{\sst{(7)}}}
\def\8{{\sst{(8)}}}
\def\sst#1{{\scriptscriptstyle #1}}

\begin{document}

\begin{center}
{\large {\bf Emergent Universe Scenario, Bouncing Universes, and Cyclic Universes in Degenerate Massive Gravity}}

\vspace{15pt}
{\large Shou-Long Li\hoch{1,2,3}, H. L\"u\hoch{3},
Hao Wei\hoch{2}, Puxun Wu\hoch{1} and Hongwei Yu\hoch{1}}

\vspace{15pt}

\hoch{1}{\it Department of Physics and Synergetic Innovation Center for Quantum Effect and Applications, Hunan Normal University, Changsha 410081, China}

\vspace{10pt}

\hoch{2}{\it School of Physics, Beijing Institute of Technology, Beijing 100081, China}

\vspace{10pt}

\hoch{3}{\it Center for Joint Quantum Studies, School of Science, Tianjin University, Tianjin 300350, China}

\vspace{40pt}

\underline{ABSTRACT}
\end{center}

We consider alternative inflationary cosmologies in massive gravity with degenerate reference metrics and study the feasibilities of the emergent universe scenario, bouncing universes, and cyclic universes. We focus on the construction of the Einstein static universe, classes of exact solutions of bouncing universes, and cyclic universes in degenerate massive gravity. We further study the stabilities of the Einstein static universe against both homogeneous and inhomogeneous scalar perturbations and give the parameters region for a stable Einstein static universe.

\vfill
 sllee\_phys@bit.edu.cn\ \ \ mrhonglu@gmail.com\ \ \ haowei@bit.edu.cn \ \ \ pxwu@hunnu.edu.cn \ \ \ hwyu@hunnu.edu.cn

\thispagestyle{empty}

\pagebreak



\newpage

\section{Introduction}

General relativity~(GR), as a classical theory describing the non-linear gravitational interaction of massless spin-2 fields, is widely accepted at the low energy limit. Nevertheless, there are still several motivations to modify GR, based on both theoretical considerations~(e.g.~\cite{Deser:1974cz, Fierz:1939ix}) and observations~(e.g.\cite{Riess:1998cb, Perlmutter:1998np}.)  One proposal, initiated by Fierz and Pauli~\cite{Fierz:1939ix}, is to assume that the mass of a graviton is nonzero. Unfortunately, the interactions for massive spin-2 fields in Fierz-Pauli massive gravity have long been thought to give rise to ghost instabilities~\cite{Boulware:1973my}. Recently, the problem has been resolved by de Rham, Gabadadze, and Tolley~(dRGT)~\cite{deRham:2010kj}, and dRGT massive gravity has attracted great attention and is studied in various areas such as cosmology~\cite{DAmico:2011eto, Comelli:2011zm, Gumrukcuoglu:2011zh, DeFelice:2012mx} and black holes~\cite{Nieuwenhuizen:2011sq, Berezhiani:2011mt}. We refer to e.g.~\cite{Hinterbichler:2011tt, deRham:2014zqa, Schmidt-May:2015vnx} and reference therein for a comprehensive introduction of massive gravity.

There are several extensions of dRGT massive gravity for different physical motivations, such as bi-gravity~\cite{Hassan:2011zd}, multi-gravity~\cite{Hinterbichler:2012cn}, minimal massive gravity~\cite{DeFelice:2015hla}, mass-varying massive gravity~\cite{Huang:2012pe}, degenerate massive gravity~\cite{Vegh:2013sk} and so on~\cite{DAmico:2012hia}. Thereinto, the degenerate massive gravity was initially proposed by Vegh~\cite{Vegh:2013sk} to study holographically a class of strongly interacting quantum field theories with broken translational symmetry. Later this theory has been studied widely in the holographic framework~\cite{Blake:2013owa, Blake:2013bqa, Andrade:2013gsa, Ge:2014aza} and black hole physics~\cite{Cai:2014znn, Xu:2015rfa, Hendi:2015hoa, Cao:2015cti, Cao:2015cza}. However, the cosmological applications of this theory are few. Recently, together with suitable cubic Einstein-Riemann gravities and some other matter fields, degenerate massive gravity was used to construct exact cosmological time crystals~\cite{Feng:2018qnx} with two jumping points, which provides a new mechanism of spontaneous time translational symmetry breaking to realize the bouncing and cyclic universes that avoid the initial spacetime singularity. It is worth noting that it is higher derivative gravity, not massive gravity, that is indispensable for the realization of cosmological time crystals, which involves discontinuity in the time derivative of the cosmological scale factor at the turning points. On the other hand, one can also consider smooth bouncing universes, and cyclic models. In the framework of Einstein gravity such models will necessarily violate the energy condition. In this paper, we consider degenerated massive gravity to study these models.       

      Actually, it is valuable to investigate alternative inflationary cosmological models within the standard big bang framework, because traditional inflationary cosmology~\cite{Guth:1980zm, Brout:1977ix, Starobinsky:1980te, Sato:1980yn} suffers from both initial singularity problem~\cite{Hawking:1969sw} and trans-Planckian problem~\cite{Martin:2000xs}. By introducing a mechanism for a bounce in cosmological evolution,  both the trans-Planckian problem and an initial singularity can be avoided. The bouncing scenario can be constructed via many approaches such as matter bounce scenario~\cite{Finelli:2001sr}, pre-big-bang model~\cite{Gasperini:1992em}, ekpyrotic model~\cite{Khoury:2001wf}, string gas cosmology~\cite{Brandenberger:1988aj}, cosmological time crystals~\cite{Feng:2018qnx} and so on~\cite{Novello:2008ra,Battefeld:2014uga, Brandenberger:2016vhg}.  The cyclic universe, e.g.~\cite{Steinhardt:2001st}, can be viewed as the extension of the bouncing universe since it brings some new insight into the original observable Universe~\cite{Lehners:2008vx}. Another direct solution to the initial singularity proposed by Ellis {\it et al.}~\cite{Ellis:2002we, Ellis:2003qz}, i.e., the emergent universe scenario, is assuming that the universe inflates from a static beginning, i.e., the Einstein static universe, and reheats in the usual way. In this scenario, the initial universe has a finite size and some past-eternal inflation, and then evolves to an inflationary era in the standard way. Both horizon problem and the initial singularity are  absent due to the initial static state. Actually, these alternative inflationary cosmologies have been studied in different class of massive gravities. The bouncing universes, and cyclic universes have been studied in mass-varying massive gravity~\cite{Cai:2012ag}. The emergent scenario has been also studied  in dRGT massive gravity~\cite{Parisi:2012cg, Zhang:2013noa} and bi-gravity~\cite{Mousavi:2016eof, Mousavi:2018cda}.  To our knowledge, these alternative inflationary models have not been studied in degenerate massive gravity.  For our purpose, we would like to study the feasibilities of an emergent universe, bouncing universes, and cyclic universes in massive gravity with degenerate reference metrics.

The remaining part of this paper is organized as follows. In Sec.~\ref{sec2}, we give a brief review of the massive gravity and its equations of motion. In Sec.~\ref{sec3}, we study the emergent universe in degenerate massive gravity with a perfect fluid. First we obtain the exact Einstein static universe solutions in several cases. Then we give the linearized equations of motion and discuss the stabilities against both homogeneous and inhomogeneous scalar perturbations. We give the parameters regions of stable Einstein static universes. In Sec.~\ref{sec4}, we construct exact solutions of the bouncing universes, and cyclic universes in degenerate massive gravity with a cosmological constant and axions. We conclude our paper in Sec.~\ref{sec5}.


\section{ Massive gravity }\label{sec2}

In this section, following e.g.~\cite{deRham:2010kj}, we briefly review massive gravity. The four-dimensional action ${\cal S}$ of massive gravity is given by
\be
{\cal S} = \frac{M_{pl}^2}{2}\int d^4 x \sqrt{-g} \left( R + m^2 ({\cal U}_2 +c_3 {\cal U}_3 +c_4 {\cal U}_4 ) \right) +{\cal S}_m  \,, \label{action}
\ee
where $M_{pl}$ is the Plank mass and we assume $M_{pl}^2/2 =1 $ in the rest discussion, ${\cal S}_m$ is the action of matters, $R$ is the Ricci scalar, $g$ represents the determinant of $g_{\mu\nu}$, $m$ represents the mass of graviton, $c_i$ are free parameters and ${\cal U}_i$ are interaction potentials which can be expressed as follows:
\be
\begin{split}
&{\cal U}_2 = [K]^2 -[K^2] \,,    \\
&{\cal U}_3 = [K]^3 -3[K] [K^2] + 2[K^3] \,, \\
&{\cal U}_4 = [K]^4 -6[K^2] [K]^2  +8[K^3] [K] + 3[K^2]^2 -6 [K^4] \,,
\end{split}
\ee
where the regular brackets denote traces such as $[K] = \textup{Tr}[K] = {K^\mu}_\mu$. ${K^\mu}_\nu$ is given by
\be
{K^\mu}_\nu = {\delta^\mu}_\nu - {W^\mu}_\nu \,,
\ee
and obeys
\be
{W^\mu}_\lambda {W^\lambda}_\nu = {(\sqrt{M})^\mu}_\lambda {(\sqrt{M})^\lambda}_\nu = {M^\mu}_\nu \,, \quad \textup{with} \quad {M^\mu}_\nu = g^{\mu\lambda} f_{\lambda\nu} \,, \label{sqrtrule}
\ee
where $f$ is a fixed symmetric tensor and  called a reference metric, which is given by
\be
f_{\mu\nu} = \partial_\mu \phi^a \partial_\nu \phi^b \eta_{ab} \,, \label{referenece}
\ee
where $\eta_{ab} = \textup{diag} (-1, 1, 1, 1)$ is the Minkowski background and  $\phi^a$ are the St\"uckelberg fields introduced to restore diffeomorphism invariance~\cite{ArkaniHamed:2002sp}.  In the limit of $m \rightarrow 0$, massive gravity reduces to GR. The equations of motion are given by
\be
R_{\mu\nu} -\frac12 g_{\mu\nu} R +m^2 X_{\mu\nu}= T_{\mu\nu} \,, \label{eom}
\ee
with
\begin{align}
\begin{split}
X_{\mu\nu} &= -\frac12 g_{\mu\nu} {\cal U}_2 +W_{\mu\nu} [K] +W_{\mu\nu}  - W_{\mu\nu}^2 +c_3 \Big(-\frac12 g_{\mu\nu} {\cal U}_3 -\frac32 W_{\mu\nu} {\cal U}_2 +3 [K] (W_{\mu\nu} \\
&\quad -W_{\mu\nu}^2) -3 (W_{\mu\nu} -2 W_{\mu\nu}^2 +W_{\mu\nu}^3)\Big) +c_4 \big[ -\frac12 g_{\mu\nu} {\cal U}_4 -2 W_{\mu\nu} {\cal U}_3 +6 \ {\cal U}_2 (W_{\mu\nu} \\
&\quad -W_{\mu\nu}^2) -12 [K] (W_{\mu\nu} -2 W_{\mu\nu}^2 +W_{\mu\nu}^3) +12 (W_{\mu\nu} -3 W_{\mu\nu}^2 +3 W_{\mu\nu}^3 - W_{\mu\nu}^4) \big] \,,
\end{split} \label{eom1}\\
W_{\mu\nu} &= g_{\mu\lambda} {W^\lambda}_\nu \,,\quad W_{\mu\nu}^2 = W_{\mu\lambda} {W^\lambda}_\nu \,,\quad W_{\mu\nu}^3 = W^2_{\mu\lambda} {W^\lambda}_\nu \,,\quad W_{\mu\nu}^4 = W^3_{\mu\lambda}  {W^\lambda}_\nu\,, \label{eom2}
\end{align}
where the energy-momentum tensor $T_{\mu\nu} = -\frac{1}{\sqrt{-g}}\frac{\partial {\cal S}_m}{\partial g^{\mu\nu}}$. We refer to e.g.~\cite{Hinterbichler:2011tt, deRham:2014zqa, Schmidt-May:2015vnx} and reference therein for more details of massive gravity.

Generally, all the St\"uckelberg fields $\phi^a$ are nonzero in massive gravity and the rank of the matrix $f$~(\ref{referenece}) is full, i.e., $\textup{rank} (f) =4$. In Ref.~\cite{Vegh:2013sk}, there are two spatial nonzero St\"uckelberg fields which break the general covariance in  massive gravity. The matrix $f$ has a rank 2 and thus, is degenerate. The massive gravity with degenerate reference metrics is called degenerate massive gravity. For our purpose, we set only the temporal St\"uckelberg field to equal to zero. It follows that the massive gravity we consider in this paper has degenerate reference metrics of rank 3.  And the unitary gauge of the corresponding St\"uckelberg fields is defined simply by $\phi^a = x^\mu \delta^a_\mu$. So $\phi^a$ are given by ~\cite{Feng:2018qnx}
\be
\phi^a = a_m (0, x, y, z) \,,
\ee
in the basis $(t, x, y, z)$, where $a_m$ is  a positive constant.


\section{Emergent universe scenario}\label{sec3}

In this section, we consider the realization of the emergent universe scenario in the context of degenerate massive gravity. We consider only a spatially flat Friedmann-Lema\^itre-Robertson-Walker~(FLRW) metric because the St\"uckelberg fields in degenerate massive gravity are chosen in a spatially flat basis. On the other hand, based on the latest astronomical observations~\cite{Bennett:2012zja, Aghanim:2018eyx}, the Universe is at good consistency with the standard spatially flat case.  In the following discussion, we assume that the matter field is composed of perfect fluids. Firstly we construct the Einstein static universe in several cases. Then we study the stability against both homogeneous and inhomogeneous scalar perturbations.


\subsection{Einstein static universe }\label{sec31}

The spatially flat FLRW metric is given by
\be
ds^2 = -dt^2 +a(t)^2 \left(dx^2 +dy^2 +dz^2\right) \,. \label{frw}
\ee
The energy-momentum tensor corresponding to perfect fluids is given by
\be
T_{\mu\nu} = (\rho + P) u_\mu u_\nu +P g_{\mu\nu} \,, \quad \textup{with} \quad P = w \rho \,, \label{fluid}
\ee
where $\rho$ and $P$ represent the energy density and pressure respectively, $w$ is the constant equation-of-state~(EOS) parameter, and velocity 4-vector $u^\mu$ is given by
\be
u^\mu = \left(1, 0, 0, 0\right) \,, \quad \textup{satisfying} \quad u^\mu u_\mu = -1 \,. \label{velocity}
\ee
Substituting Eqs.~(\ref{frw}) and (\ref{fluid}) into the equations of motion~(\ref{eom}), the Friedmann equations are given by
\begin{align}
\begin{split}
&\dot{a}^2 + \left(2(2 c_3 +2 c_4 +1) m^2 -\frac{\rho}{3}\right) a^2 -3  m^2 a_m (3 c_3 +4 c_4 +1) a -\frac{ (c_3 +4 c_4) m^2 a_m^3}{a} \\
& + (6 c_3 +12 c_4 +1) m^2 a_m^2 =0 \,,
\end{split} \label{friedmann} \\
\begin{split}
& \ddot{a} + \frac{\dot{a}^2}{2 a} + \left( 3 (2 c_3 +2 c_4 +1) m^2 +\frac{w \rho }{2} \right) a +\frac{(6 c_3 +12 c_4 +7) m^2 a_m^2}{2 a} -(9 c_3 +12 c_4 \\
&+7) m^2 a_m =0 \,, \label{friedmann2}
\end{split}
\end{align}
where the dot denotes the derivative with respect to time.
For the sake of obtaining the Einstein static universe,  we let the scale factor $a(t)=a_0 = \textup{const.} \ne 0$ and $\dot{a} =\ddot{a} =0$. We request $a_0 <a_m$~\cite{DeFelice:2012mx} to avoid the ghost excitation from massive gravity. The energy density $\rho$ can be solved from the Friedmann equation~(\ref{friedmann}),
\be
\rho = \frac{3 m^2 (n-1) \left(c_3 \left(4 n^2-5 n+1\right)+4 c_4 (n-1)^2+n (2 n-1)\right)}{n^3} >0 \,, \label{rhocond}
\ee
where
\be
n = \frac{a_0}{a_m} \,, \quad \textup{with}  \quad 0< n <1 \,. \label{ncond}
\ee
Substituting Eqs.~(\ref{rhocond}) and (\ref{ncond}) into (\ref{friedmann2}), the final independent equation is given by
\be
e_3 n^3 +e_2 n^2 +e_1 n +e_0 =0 \,, \label{final}
\ee
with
\be
\begin{split}
e_0 &= -3 (c_3+4 c_4) w \,, \quad e_1 = 7+3 w +6 (c_3 +2 c_4) (1+3 w) \,,  \\
e_2 &= -14 -9 w -3 (3 c_3 +4 c_4) (2+3 w) \,, \quad e_3 = 6 (1+2 c_3 +2 c_4) (1+w) \,.
\end{split}
\ee
The Einstein static universe solution is given by $a_0 = n\ a_m$.  Because there are several parameters in the Eq.~(\ref{final}), we will discuss them in different cases.


\subsubsection{Case 1: $e_3 =e_2 =0$, $e_1 \ne 0$, $e_0 \ne 0$ }\label{sec311}
In this case, Eq.~(\ref{final}) reduces to a simple linear equation. The Einstein static solution is given by
\be
n = -\frac{e_0}{e_1} \,. \label{case1}
\ee
Note that the reality conditions~(\ref{rhocond}) and (\ref{ncond}) are required. We find that the Einstein static universe~(\ref{case1}) can  exist in the following two cases:

Case (1.1):   For $  c_4 = -1/2 -c_3 \,, -1/9< c_3<1/3 \,, w= -(6 c_3-2 )/(9 c_3+9 )\,,$  the solution is given by
 \be
n = \frac{2 \left(3 c_3-1\right) \left(3 c_3+2\right)}{18 c_3^2+3 c_3-7} \,, \quad \rho = \frac{9 \left(c_3+1\right) \left(9 c_3+1\right) \left(18 c_3^2+3 c_3-7\right) m^2}{8 \left(3 c_3-1\right){}^3 \left(3 c_3+2\right){}^2} \,. \label{sol11}
\ee

Case (1.2):   For $ 13/27 < c_3 <  5/6  \,, c_4 = -(9 c_3 -5)/12 \,,  w= -1$, the solution is given by
\be
n = \frac{6 c_3-5}{6 \left(c_3-1\right)} \,, \quad \rho = \frac{\left(27 c_3-13\right) m^2}{\left(5-6 c_3\right){}^2} \,.	\label{sol12}
\ee


\subsubsection{Case 2: $e_3 =0$, $e_2 \ne 0$ }\label{sec312}
In this case, Eq.~(\ref{final}) reduces to a quadratic equation. The Einstein static solutions are given by
\be
n = n_{\pm} = \frac{-e_1 \pm \sqrt{e_1^2 -4 e_0 e_2}}{2 e_2} \,, \quad \textup{with} \quad e_1^2 -4 e_0 e_2\ge 0 \,.
\ee
We discuss the existence of the two solutions respectively. Both cases require reality conditions~(\ref{rhocond}) and (\ref{ncond}). The existence of $n = n_+$ requires the following two cases:

Case (2.1):  For  $c_4 = -1/2 -c_3 $, and
\be
 \left\{
\begin{aligned}
 & -\frac23 < c_3 < -\frac19 \,, \quad  w < 0 \,, \\
 &  c_3 = -\frac19 \,, \quad   -\frac59 <w < 0  \,, \\
 &  -\frac19 < c_3 < \frac16 \,, \quad   -1-4 c_3 +\frac{2}{3} \sqrt{27 c_3^2 +21 c_3 +2 } \le w <0  \,,
\end{aligned}
\right.
\ee
the solution is given by
\be
\begin{split}
 n &= n_{+} =\frac{15 w -1 +6 c_3 (1 +3 w) + \sqrt{36 c_3^2 -12 c_3 +1 +18 w +72 w c_3 +9 w^2 }}{2 (9 w -2 + 3 c_3 (3 w  +2 ))} \,, \\
 \rho &=  \frac{3 m^2 (1 -n_+ ) \left( 2+3 c_3 -3 n_+ -3 c_3 n_+ \right)}{n_{+}^3} \,.
\end{split} \label{sol21}
\ee

Case (2.2):  For  $w = -1 $, and
\be
 \left\{
\begin{aligned}
 & -\frac73 < c_3 \le \frac{1}{15} \,, \quad \frac{27 c_3^2 +38 c_3 -21}{64} < c_4 < -\frac{c_3}{4} \,, \\
 &  \frac{1}{15} < c_3 < \frac{2}{3} \,, \quad -\frac{9 c_3^2 -9 c_3 +4}{12} \le c_4 < -\frac{c_3}{4}  \,,
\end{aligned}
\right.
\ee
the solution is given by
\be
\begin{split}
 n &= n_{+} =\frac{12 c_3+24 c_4 +5 - \sqrt{ 9 c_3^2 +51 c_3 +25 -36 c_4 -144 c_3 c_4 -144 c_4^2 }}{23 +45 c_3 +60 c_4 } \,, \\
 \rho &=  \frac{3 m^2 (n_+ -1 ) \left(  c_3 +4 c_4 -n_+ -5 c_3 n_+ -8 c_4 n_+ +2 n_+^2 +4 c_3 n_+^2 +4 c_4 n_+^2 \right)}{n_{+}^3} \,.
\end{split} \label{sol22}
\ee
The existence of $n = n_-$ requires the following two cases:

Case (2.3):   For  $c_4 = -1/ 2 -c_3 $, and
\be
 \left\{
\begin{aligned}
 & -\frac19 < c_3 < \frac16 \,, \quad w\ge -1-4 c_3 +\frac{2}{3} \sqrt{27 c_3^2 +21 c_3 +2 } \,,\quad \textup{and}\quad  w \ne -\frac{2 \left(3 c_3-1\right)}{9 \left(c_3+1\right)}  \,, \\
 &   \frac16 \le c_3 < \frac13 \,, \quad  w > 0  \,,\quad \textup{and}\quad  w \ne -\frac{2 \left(3 c_3-1\right)}{9 \left(c_3+1\right)} \,, \\
 &  c_3 \ge \frac13 \,, \quad    w > 0   \,,
\end{aligned}
\right.
\ee
the solution is given by
\be
\begin{split}
 n &= n_{-}= \frac{6 c_3 (3 w+1)+15 w-1 - \sqrt{12 (6 w-1) c_3 +36 c_3^2+9 w^2+18 w+1}}{2 \left( (9 w+6) c_3 +9 w-2\right)}  \,,  \\
 \rho &=  \frac{3 m^2 (n_{-}-1) \left(3 c_3 (n_{-}-1)+3 n_{-} -2\right)}{n_{-}^3} \,.
\end{split} \label{sol23}
\ee

Case (2.4):   For  $w = -1 $, and
\be
 \left\{
\begin{aligned}
 & \frac{1}{15} < c_3 \le \frac{13}{27} \,, \quad  -\frac{9 c_3^2 -9 c_3 +4}{12} \le c_4 < \frac{27 c_3^2 +38 c_3 -21}{64} \,, \\
 &    \frac{13}{27} < c_3 < \frac{2}{3} \,, \quad -\frac{9 c_3^2 -9 c_3 +4}{12} \le c_4 < \frac{27 c_3^2 +38 c_3 -21}{64}   \,,\quad \textup{and}\quad  c_4 \ne \frac{5-9 c_3}{12} \,, \\
 &   \frac{2}{3} \le c_3 \le \frac{5}{6} \,, \quad -\frac{ c_3 }{4} < c_4 < \frac{27 c_3^2 +38 c_3 -21}{64}   \,,\quad \textup{and}\quad  c_4 \ne \frac{5-9 c_3}{12}  \,, \\
 &  c_3 > \frac{5}{6} \,, \quad -\frac{ c_3 }{4} < c_4 < \frac{27 c_3^2 +38 c_3 -21}{64}   \,,
\end{aligned}
\right.
\ee
the solution is given by
\be
\begin{split}
 n &= n_{-}= \frac{12 c_3+24 c_4 +5 - \sqrt{ 9 c_3^2 +51 c_3 +25 -36 c_4 -144 c_3 c_4 -144 c_4^2 }}{23 +45 c_3 +60 c_4 }  \,,  \\
 \rho &=  \frac{3 m^2 (n_- -1 ) \left(  c_3 +4 c_4 -n_- -5 c_3 n_- -8 c_4 n_- +2 n_-^2 +4 c_3 n_-^2 +4 c_4 n_-^2 \right)}{n_{-}^3} \,.
\end{split} \label{sol24}
\ee


\subsubsection{Case 3: $e_3 \ne 0$ }\label{sec313}
\begin{figure}[ht]
\centerline{
\includegraphics[width=0.3\linewidth]{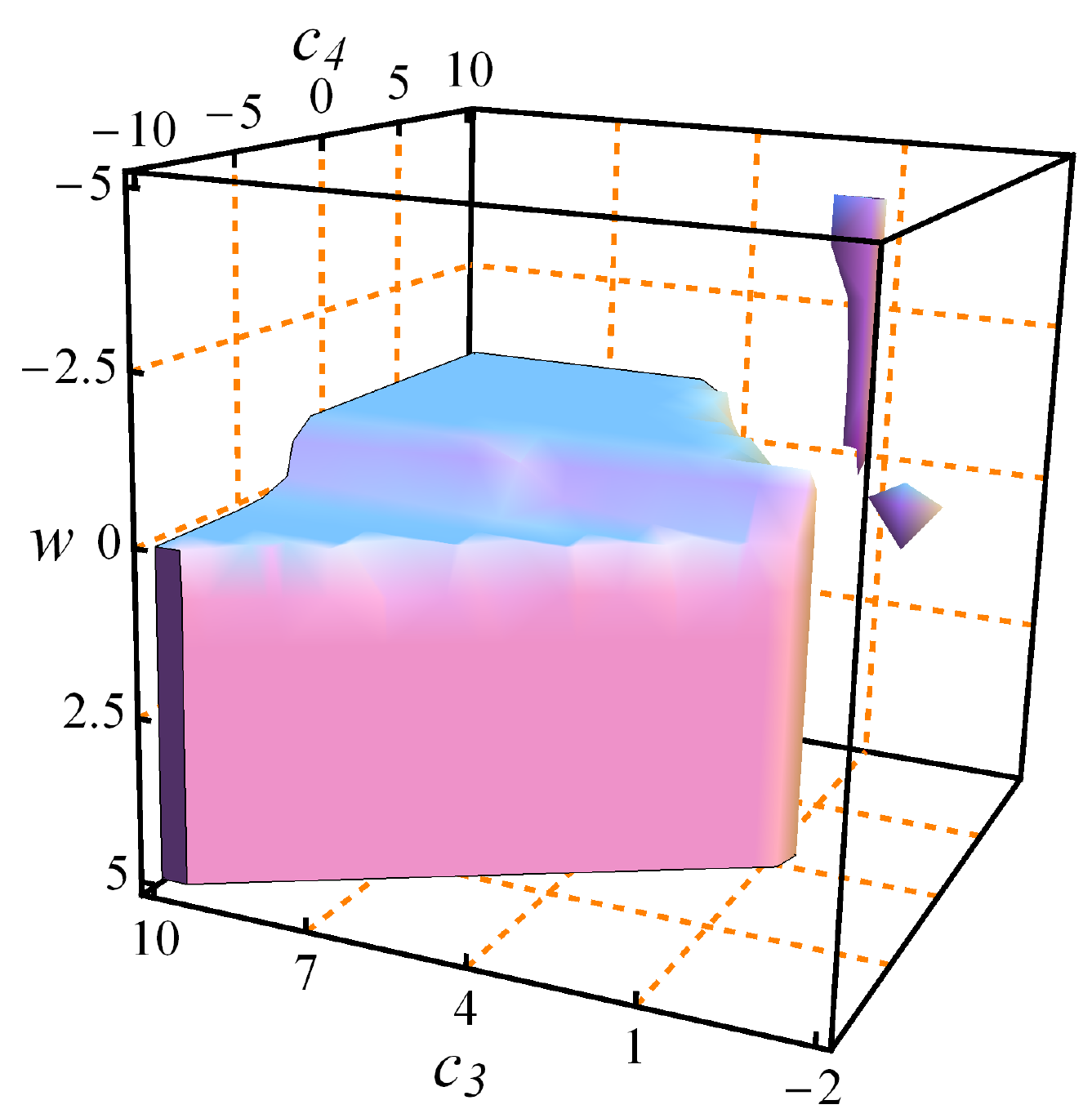}\ \ \ \ \ \includegraphics[width=0.3\linewidth]{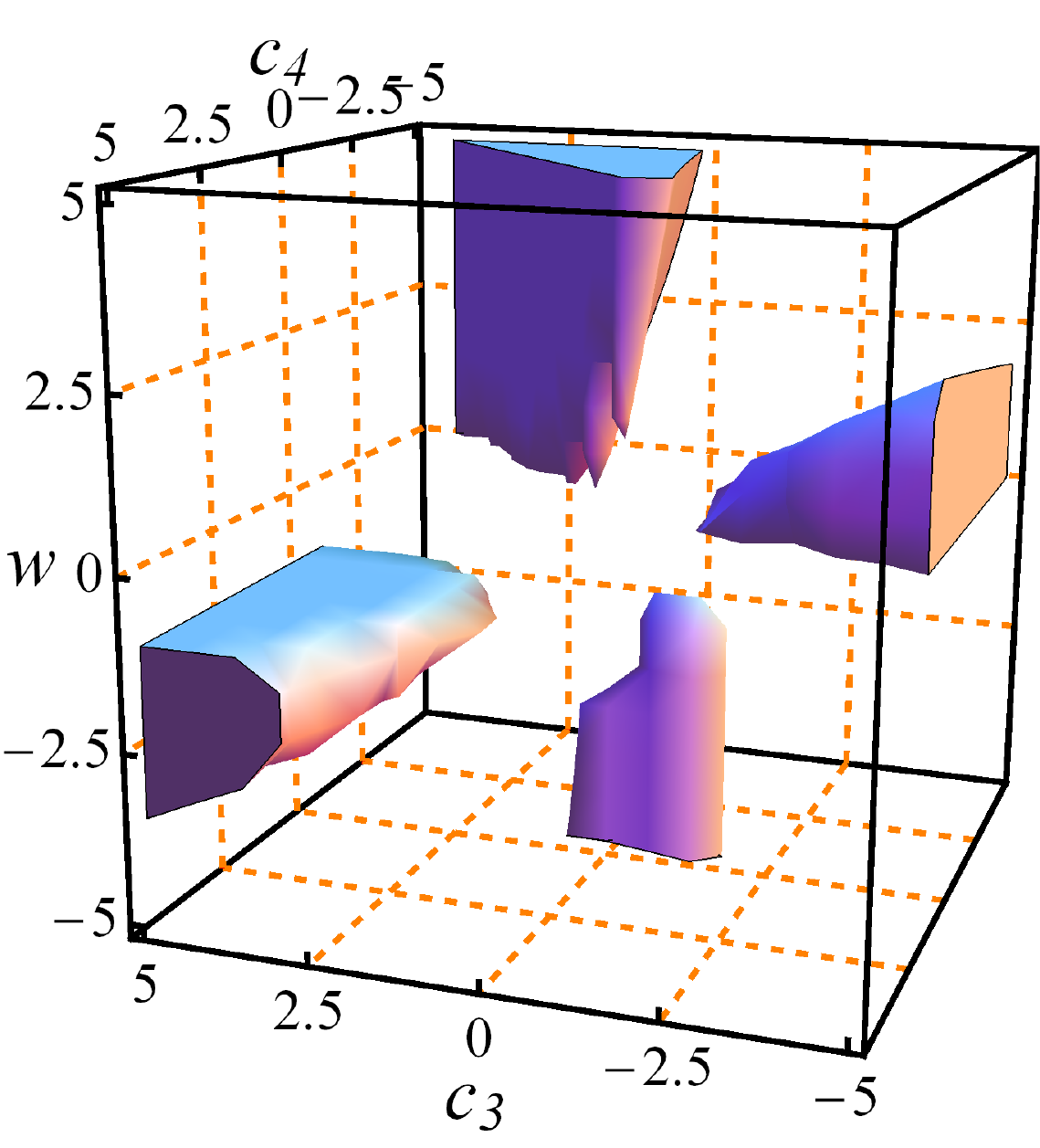}\ \ \ \ \ \includegraphics[width=0.3\linewidth]{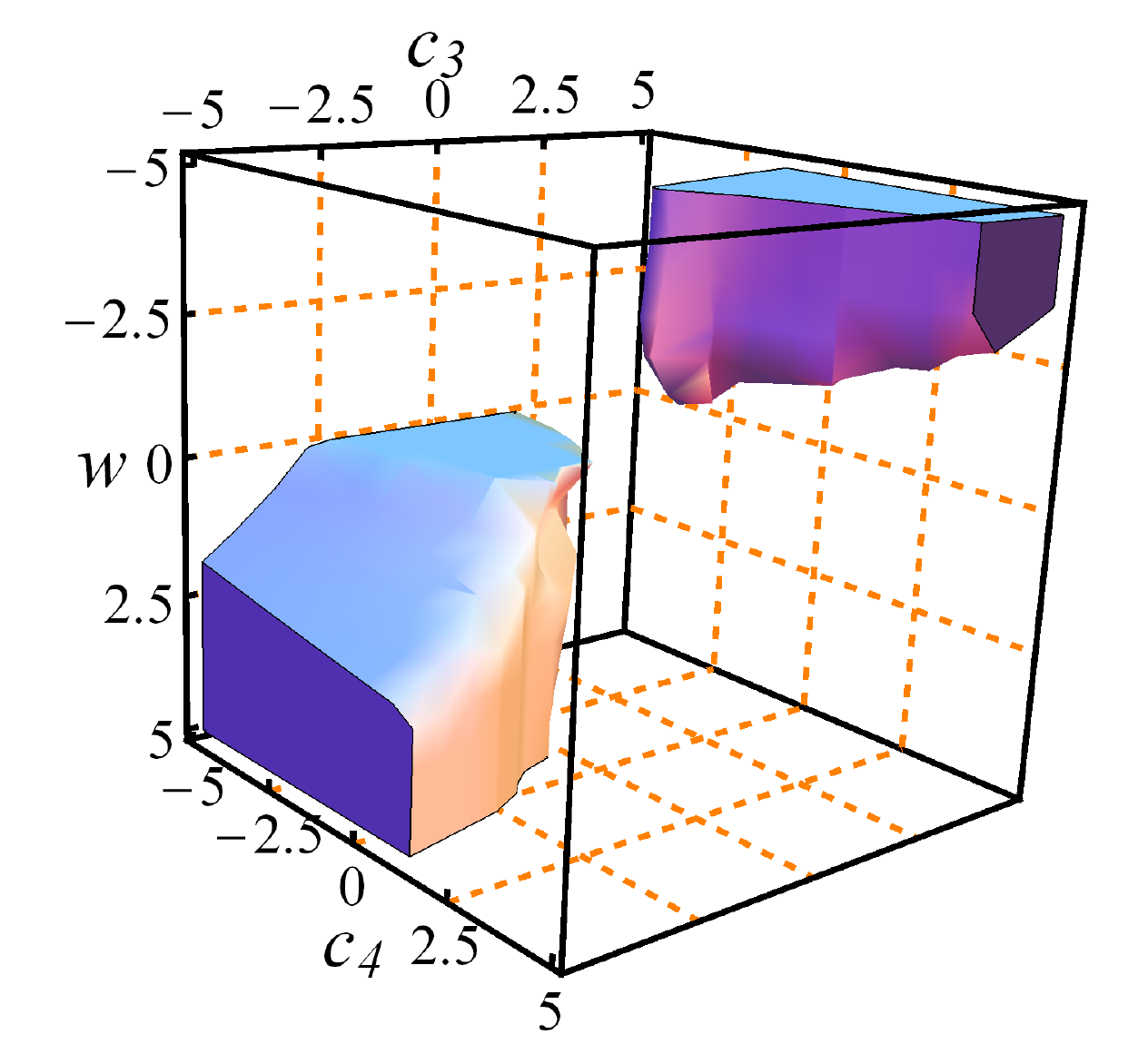}}
\caption{\small The parameters $(c_3, c_4, w)$ regions of the existence of Einstein static solutions ${n}_2$ (left), ${n}_3$ (middle) and $ {n}_4$ (right). We set $m=1$ for simplicity.}
\label{solfig}
\end{figure}

In this case, Eq.~(\ref{final}) can be rewritten as
\be
\hat{n}^3 + \hat e_1 \hat n +\hat e_0 =0 \,,
\ee
where
\be
\hat n = n + \frac{e_2}{3 e_3}  \,,\quad  \hat e_1 = -\frac{ e_2^2 -3 e_1 e_3 }{3 e_3^2} \,,\quad \hat e_0 = \frac{2 e_2^3 -9 e_1 e_2 e_3 +27 e_0 e_3^2 }{27 e_3^3} \,.\label{nsol}
\ee
For $4 \hat e_1^3 +27 \hat e_0^2 \le 0 $  and $ \hat{e}_1 <0 $, there are three real solutions which are given by
\be
\hat n = \hat n_{k+1} = 2 \sqrt{-\frac{\hat e_1}{3}} \cos \Big( \frac13 \arccos \Big (\frac{3 \hat e_0}{2 \hat e_1} \sqrt{-\frac{3}{\hat e_1}} \Big) -\frac{2 k \pi }{3} \Big)  \,,\quad k= 0, 1, 2.
\ee
For $4 \hat e_1^3 +27 \hat e_0^2 > 0$ and $ \hat e_1 <0 $, there is one real solution which is given by
\be
\hat n = \hat n_4 =  - 2 \frac{{|\hat e_0|}}{\hat e_0} \sqrt{ -\frac{\hat e_1}{3}} \cosh \Big(\frac13 \textup{arcosh} \Big(-\frac{3|\hat e_0|}{2 \hat e_1} \sqrt{-\frac{3}{\hat e_1}} \Big ) \Big )  \,.
\ee
For $\hat e_1 >0 $, there is one real solution which is given by
\be
\hat n = \hat n_5 =  - 2 \sqrt{ \frac{\hat e_1}{3}} \sinh \Big(\frac13 \textup{arsinh} \Big( \frac{3 \hat e_0}{2 \hat e_1} \sqrt{ \frac{3}{\hat e_1}} \Big ) \Big )  \,.
\ee
For $\hat e_1 =0 $, there is one real solution which is given by
\be
\hat n = \hat n_6 =  (-\hat e_0)^{\frac13} \,.
\ee
Substituting the solutions into Eqs.~(\ref{nsol}) and (\ref{rhocond}), the solutions and  energy density  are given by
\be
\begin{split}
n &= \hat{n} + \frac{14 +9w +3(3 c_3 +4 c_4)(2+3w)}{18(1+2c_3 +2c_4)(1+w)} \,,\\
\rho &= 6 \left(2 c_3+2 c_4+1\right) m^2 (9 c_3 (4 \hat{n} (w+1)-w-2)+12 c_4 (3 \hat{n} (w+1)-1)+18 \hat{n} w \\
&\quad +18 \hat{n}-9 w-4) (81 c_3^2 \left(16 \hat{n}^2 (w+1)^2+4 \hat{n} \left(w^2-1\right)-2 w^2-5 w-2\right) \\
&\quad +12 c_4 \left(108 \hat{n}^2 (w+1)^2+3 \hat{n} \left(9 w^2+16 w+7\right)+27 w^2+45 w+8\right) \\
&\quad +18 c_3 (6 c_4 \left(24 \hat{n}^2 (w+1)^2+\hat{n} \left(3 w^2-8 w-11\right)-w+1\right)+72 \hat{n}^2 (w+1)^2 \\
&\quad +\hat{n} \left(27 w^2+56 w+29\right)-2 w-7 )+144 c_4^2 (1-3 \hat{n} (w+1))^2+324 \hat{n}^2 w^2 \\
&\quad +648 \hat{n}^2 w+324 \hat{n}^2+162 \hat{n} w^2+504 \hat{n} w+342 \hat{n}+45 w+70 )/(9 c_3 (4 \hat{n} (w+1) \\
&\quad +3 w+2)+12 c_4 (3 \hat{n} (w+1)+3 w+2)+18 \hat{n} w+18 \hat{n}+9 w+14 )^3 \,.
\end{split}
\ee
There are three free parameters $c_3, c_4,$ and $w$ in the solutions. It is hard to analyze the parameters region of existence of all six solutions analytically. Instead we analyze the existence regions numerically and plot the parameters $(c_3, c_4, w)$ regions of the existence  of all solutions in Fig.~\ref{solfig}.
We find that the solutions ${n}_1, n_5$ and $ {n}_6$ cannot exist.


\subsection{Stabilities }\label{sec32}
In the previous subsection, we study the existence of the Einstein static universe in  massive gravity with degenerate reference metrics.  However, the emergent scenario does not thoroughly solve the issue of big bang singularity when perturbations are considered. For example, although the Einstein static universe is stable against small inhomogeneous perturbations in some cases~\cite{Gibbons:1987jt, Gibbons:1988bm,Mulryne:2005ef, Barrow:2003ni}, the instability exists in previous parameters range against homogeneous perturbations~\cite{Eddington:1930zz}. So it is valuable to explore the viable Einstein static universe by considering both homogeneous and inhomogeneous scalar perturbations. Actually, the stabilities of the Einstein static universe has been studied in various modified gravities, for examples, loop quantum cosmology~\cite{Parisi:2007kv}, $f(R)$ theory~\cite{Barrow:1983rx, Boehmer:2007tr, Seahra:2009ft, Goheer:2008tn}, $f(T)$ theory~\cite{Wu:2011xa, Li:2013xea},  modified Gauss-Bonnet gravity~\cite{Bohmer:2009fc, Huang:2015kca}, Brans-Dicke theory~\cite{delCampo:2007mp, delCampo:2009kp, Huang:2014fia, Miao:2016obc,Labrana:2018bkw}, Horava-Lifshitz theory~\cite{Wu:2009ah, Boehmer:2009yz, Heydarzade:2015hra},  brane world scenario~\cite{Gergely:2001tn, Zhang:2010qwa, Atazadeh:2014xsa}, Einstein-Cartan theory~\cite{Atazadeh:2014ysa}, $f(R, T)$ gravity~\cite{Shabani:2016dhj}, Eddingtong-inspired Born-Infeld theory~\cite{Li:2017ttl}, Horndeski theory~\cite{Atazadeh:2015zma, Huang:2018kqr}, hybrid metric-Palatini gravity~\cite{Boehmer:2013oxa} and so on~\cite{Clifton:2005at, Boehmer:2003iv, Goswami:2008fs, Canonico:2010fd, Zhang:2016obw, Carneiro:2009et, Boehmer:2015ina, Atazadeh:2016yeh,Khodadi:2016gyw,Khodadi:2015tsa,Bag:2014tta}. We refer to e.g.~\cite{Barrow:2003ni} and reference therein for more details of stability of the Einstein static universe. In the following discussions, we would  consider the stabilities of the Einstein static universe against both homogeneous and inhomogeneous scalar perturbations in degenerate massive gravity.


\subsubsection{Linearized Massive Gravity}\label{sec321}

Now we study the linear massive gravity with degenerate reference metrics. We use the symbols bar and tilde representing the background and the perturbation components of  the metric respectively. First, we obtain the linearized equations of motion The perturbed metric can be written as
\be
g_{\mu\nu} = \bar{g}_{\mu\nu} + h_{\mu\nu} \,, \label{pertg}
\ee
where $\bar{g}_{\mu\nu}$ is the background metric which is given by~Eq.~(\ref{frw}) with $a= a_0$ and  $h_{\mu\nu}$ is a small perturbation. For our purpose, we consider scalar perturbations in the Newtonian gauge. ${h_\mu}^\nu$ is given by
\be
{h_\mu}^\nu = \textup{diag} \left(-2 \Psi \,, 2 \Phi \,, 2 \Phi \,, 2 \Phi\right) \,, \label{hdu}
\ee
where $\Psi$ and $\Phi$ are functions of $(t, x, y, z)$. For scalar perturbations, it is useful to perform a harmonic decomposition~\cite{Harrison:1967zza},
 Now the indexes are lowered and raised by the background metric unless otherwise stated. By using the relation $g^{\mu\nu} g_{\nu\lambda} ={\delta^\mu}_\lambda$, the inverse metric is perturbed by
\be
\widetilde{g}^{\mu\nu} = -\bar{g}^{\mu\rho} \bar{g}^{\nu\sigma} h_{\rho\sigma} \,.
\ee
So the perturbed $M$ can also be  written as
\be
 \widetilde{M} {^\mu}_\nu = \tilde{g}^{\mu\lambda} \partial_\lambda \phi^a \partial_\nu \phi^b \eta_{ab} \,.
\ee
According to Eq.~(\ref{sqrtrule}), we have $M {^\mu}_\nu = (\bar{W} {^\mu}_\lambda+ \widetilde{W} {^\mu}_\lambda ) (\bar{W} {^\lambda}_\nu+ \widetilde{W} {^\lambda}_\nu )$, i.e.,
\be
 \bar{M}  {^\mu}_\nu + \widetilde{M}  {^\mu}_\nu = \bar{W} {^\mu}_\lambda \bar{W}{^\lambda}_\nu + \bar{W} {^\mu}_\lambda \widetilde{W}{^\lambda}_\nu + \widetilde{W} {^\mu}_\lambda \bar{W} {^\lambda}_\nu \,.
\ee
So we have
\be\begin{split}
&\bar{W} {^\mu}_\nu = \sqrt{\bar{M} } {^\mu}_\nu = \textup{diag} \big(0, n^{-1}, n^{-1}, n^{-1} \big) \,,  \\
&\widetilde{W} {^0}_0  = 0  \,, \quad \widetilde{W} {^i}_i  = \frac{\widetilde{M} {^i}_i }{2 \bar{W} {^i}_i}    \,, \quad \widetilde{W} {^0}_i  =  \widetilde{W} {^i}_0  =  0  \,,
\end{split}\ee
where ``\,0\," and ``\,$i, j$\," denote time and space components respectively, and the same index does not mean the Einstein rule. For perfect fluids, the perturbations of energy density and pressure are $\widetilde{\rho}$ and $\widetilde{P} = w \widetilde{\rho}$ respectively. The perturbations of velocity are given by
\be
  \widetilde{u}_0 = \widetilde{u}^0 = \frac{h_{00}}{2}   \,, \quad \widetilde{u}^i = \bar{g}^{ij} \, \widetilde{u}_j = \bar{g}^{ij} \bar{\nabla}_j U  \,, \label{pertu}
\ee
where  $\tilde \rho$ and $\widetilde U$ are also functions of $(t, x, y, z)$. The perturbed energy momentum tensor is given by
\be
\widetilde{T}_{\mu\nu} = P_0 \, \widetilde{g}_{\mu\nu} +\widetilde{P} \, \bar{g}_{\mu\nu} +  (\widetilde{\rho} + \widetilde{P}) u_\mu u_\nu +(\rho_0 + P_0) \widetilde{u}_\mu u_\nu +(\rho_0 + P_0) u_\mu \widetilde{u}_\nu  \,, \label{pertT}
\ee
where $u^\mu$ represents the background components and is given by Eq.~(\ref{velocity}). Considering above expressions, the linearized equations of Eqs.~(\ref{eom})-(\ref{eom2}) are given by
\be
\widetilde{R}_{\mu\nu} -\frac12 \bar{g}_{\mu\nu} \widetilde{R} -\frac12 \tilde{g}_{\mu\nu} \bar R + m^2 \widetilde{X}_{\mu\nu} = \widetilde T_{\mu\nu} \,, \label{lineareom}
\ee
where
\begin{align}
\begin{split}
\widetilde X_{\mu\nu} &= -\frac12 \widetilde g_{\mu\nu} {\cal \bar U}_2 -\frac12 \bar g_{\mu\nu} {\cal \widetilde U}_2 + \widetilde W_{\mu\nu} [\bar K] + \bar W_{\mu\nu}  \widetilde{[K]} + \widetilde W_{\mu\nu}  - \widetilde W_{\mu\nu}^2 +c_3 \big[-\frac12 \widetilde g_{\mu\nu} {\cal \bar U}_3  \\
&\quad  -\frac12 \bar g_{\mu\nu} {\cal \widetilde U}_3 -\frac32 \widetilde W_{\mu\nu} {\cal \bar U}_2 -\frac32 \bar W_{\mu\nu} {\cal \widetilde U}_2 +3 \widetilde{[K]} (\bar W_{\mu\nu} -\bar W_{\mu\nu}^2)+3 {[\bar K]} (\widetilde W_{\mu\nu} -\widetilde W_{\mu\nu}^2)  \\
&\quad  -3 (\widetilde W_{\mu\nu} -2 \widetilde W_{\mu\nu}^2 +\widetilde W_{\mu\nu}^3)\big] +c_4 \big[ -\frac12 \widetilde g_{\mu\nu} {\cal \bar U}_4 -\frac12 \bar g_{\mu\nu} {\cal \widetilde U}_4 -2 \widetilde W_{\mu\nu} {\cal \bar U}_3 -2 \bar W_{\mu\nu} {\cal \widetilde U}_3  \\
&\quad +6 \ {\cal \widetilde U}_2 (\bar W_{\mu\nu} -\bar W_{\mu\nu}^2)+6 \ {\cal \bar U}_2 (\widetilde W_{\mu\nu} -\widetilde W_{\mu\nu}^2) -12 \widetilde{[K]} (\bar W_{\mu\nu}-2\bar W_{\mu\nu}^2 +\bar W_{\mu\nu}^3)   \\
&\quad  -12 [\bar K] (\widetilde W_{\mu\nu}-2 \widetilde W_{\mu\nu}^2 +\widetilde W_{\mu\nu}^3)+12 (\widetilde W_{\mu\nu} -3 \widetilde W_{\mu\nu}^2 +3 \widetilde W_{\mu\nu}^3 -\widetilde W_{\mu\nu}^4) \big] \,,
\end{split} \\
\begin{split}
 \widetilde W_{\mu\nu} &= \bar g_{\mu\lambda} \widetilde{W}{^\lambda}_\nu + \widetilde g_{\mu\lambda} \bar{W}{^\lambda}_\nu \,,\quad  \widetilde W_{\mu\nu}^2 =  \widetilde W_{\mu\lambda} \bar {W}{^\lambda}_\nu +\bar W_{\mu\lambda} \widetilde {W}{^\lambda}_\nu \,,\\
  \widetilde {W}_{\mu\nu}^3 &=  \widetilde W_{\mu\lambda}^2 \bar{W}{^\lambda}_\nu + \bar W_{\mu\lambda}^2 \widetilde{W}{^\lambda}_\nu \,, \quad  \widetilde W_{\mu\nu}^4 =  \widetilde W_{\mu\lambda}^3 \bar{W}{^\lambda}_\nu + \bar W_{\mu\lambda}^3 \widetilde{W}{^\lambda}_\nu \,,
\end{split} \\
\begin{split}
\widetilde{[K] } &=  -\widetilde{W} {^\mu}_\mu  \,,\quad  \widetilde{[K^2] } = -2 \bar{K} {^\mu}_\nu \widetilde{W} {^\nu}_\mu \,, \\
\widetilde{[K^3] } &= -3 \bar{K} {^\mu}_\nu \bar{K} {^\nu}_\lambda \widetilde{W} {^\lambda}_\mu \,, \quad \widetilde{[K^4] } = -4 \bar{K} {^\mu}_\nu \bar{K} {^\nu}_\lambda \bar{K} {^\lambda}_\rho \widetilde{W} {^\rho}_\mu \,,
\end{split} \\
\begin{split}
{\cal \widetilde U}_2 &= 2 [\bar K] \widetilde{[K] } -\widetilde{[K^2] } \,,\quad {\cal \widetilde U}_3 = 3 [\bar K]^2 \widetilde{[K] } -3  [\bar K] \widetilde{[K^2 ] } -3  [\bar K^2] \widetilde{[K ] } +2 \widetilde{[K^3] } \,,\\
{\cal \widetilde U}_4 &= 4 [\bar K]^3 \widetilde{[K] } -6 [\bar K]^2 \widetilde{[K^2] } -12 [\bar K] [\bar K^2] \widetilde{[K] } + 8 [\bar K] \widetilde{[K^3] }+ 8 [\bar K^3] \widetilde{[K]} +6 [\bar K^2] \widetilde{[K^2] } \\
&\quad -6 \widetilde{[K^4] } \,.
\end{split}
\end{align}
It is useful to perform a harmonic decomposition of $(\Psi, \Phi, \tilde \rho, U)$,
\be
\begin{split}
\Psi = \Psi_k(t)\, Y_k (x, y, z) \,,\quad \Phi = \Phi_k(t) \,Y_k (x, y, z) \,,\\
\widetilde{\rho} = \rho \, \xi_k(t) \,Y_k (x, y, z) \,,\quad U = U_k(t) \,Y_k (x, y, z) \,.
\end{split}  \label{decomposition}
\ee
In these expressions, summation over co-moving wavenumber $k$  are implied. The harmonic function $Y_k (x, y, z)$ satisfies~\cite{Harrison:1967zza}
\be
\Delta Y_k (x, y, z) =  -\kappa^2 Y_k (x, y, z) \,, \label{laplacian}
\ee
where $\Delta$ is Laplacian operator and $\kappa$ is separation constant. For spatially flat universe, we have $\kappa^2 =\ell^2$ where the modes are discrete $(\ell= 0, 1, 2\dots)$~\cite{Barrow:2003ni, Huang:2015kca}. Substituting Eqs.~(\ref{hdu}) and (\ref{decomposition}) into (\ref{lineareom}), after some algebra, we find
\be
\begin{split}
\Phi_k(t) &=  \Psi_k (t) \,,\quad U_k =\frac{2 \dot{\Psi}_k(t)}{\rho_0 (1+w)} \,, \\
\xi_k(t) &= (2 \kappa^2 n  +a_m^2 (12 c_4 m^2 (1-n)^2 (5-2 n) + 3 c_3 m^2 (5-24 n+27 n^2 -8 n^3) \\
&\quad  +n( -3 m^2 (4 -9 n +4 n^2) +2 n^2 \rho)) ) \Psi_k (t) /(n^3 a_m^2 \rho_0 ) \,,
\end{split}
\ee
where $\Psi_k(t)$  satisfies a second order ordinary differential equation
\be
\ddot{\Psi}_k +  Z \Psi_k =0 \,,\label{ode}
\ee
with
\be
\begin{split}
Z &= (2 \kappa^2 n w +a_m^2 ( 4 n^3 \rho w -m^2 (12 (1+2 c_3 +2 c_4) (w -1) n^3 +(14 +18 c_3 +24 c_4  \\
&\quad -27 w -81 c_3 w -108 c_4 w ) n^2  +12 (1+6 c_3 +12 c_4) w n -15 (c3+4 c_4) w ))/(2 n^3 a_m^2) \,.\label{Zdef}
\end{split}
\ee
To analyze the stabilities of  the Einstein static universe in massive gravity with a degenerate reference metric, we require the condition of the existence of the oscillating solution of Eq.~(\ref{ode}) which is given by
\be
Z>0  \,.\label{Zcond}
\ee
In the following discussions, we study the parameters region satisfying reality conditions~(\ref{rhocond}) and (\ref{ncond}), and the stability condition~(\ref{Zcond}) for the Einstein static flat universes against both homogeneous and inhomogeneous perturbations in different cases.


\subsubsection{Case 1: $e_3 =e_2 =0$, $e_1 \ne 0$, $e_0 \ne 0$ }\label{sec322}
The stabilities of the Einstein static universe~(\ref{sol11}) require
\be
c_4 = -\frac12 -c_3 \,,\quad -\frac{1}{9 }< c_3 < \frac13  \,, \quad w =\frac{2 (1 -3 c_3)}{9 (1+c_3)} \,,\quad a_m^2 < \frac{4 (3 c_3 -1) \kappa^2}{3 m^2  ( 18 c_3^2 +3 c_3 -7)}  \,.
\ee
 It is easy to see that the Einstein static flat universe in degenerate massive gravity can be stable under inhomogeneous scalar perturbations ($\kappa^2>0$), but not be stable against homogeneous scalar perturbations ($\kappa^2 =0$).  There is no stable region for an Einstein static universe~(\ref{sol12}) under either homogeneous or inhomogeneous scalar perturbations.


\subsubsection{Case 2: $e_3=0$, $e_2 \ne 0$}\label{sec323}

In the case (2.1), apart from existence conditions, the stabilities of the Einstein static universe~(\ref{sol21}) require another condition
\be
0 \le \kappa^2 < -\frac{ m^2 a_m^2}{2 w} \sqrt{ 36 c_3^2 -12 c_3  +72 w c_3 +9 w^2 +18 w +1} \,.\label{cond21}
\ee
 In the case (2.2), apart from existence conditions, the stabilities of the Einstein static universe~(\ref{sol22}) require another condition
\be
0 \le \kappa^2 <  m^2 a_m^2 \sqrt{ 9 c_3^2 -9 c_3  +12 c_4 +4} \,.
\ee
In the case (2.3), the stabilities of  the Einstein static universe~(\ref{sol23}) require conditions
\be
\begin{split}
&-\frac19 < c_3 < \frac16 \,, \quad -1-4 c_3 +\frac{2}{3} \sqrt{27 c_3^2 +21 c_3 +2 } < w < 0   \,,\quad   \textup{and} \quad \\
& 0 \le \kappa^2 < -\frac{ m^2 a_m^2}{2 w} \sqrt{ 36 c_3^2 -12 c_3  +72 w c_3 +9 w^2 +18 w +1} \,,\\
\end{split} \label{cond231}
\ee
and
\be
 \left\{
\begin{aligned}
 & -\frac19 < c_3 < \frac16 \,, \quad w >0 \,,\quad \textup{and}\quad  w \ne -\frac{2 \left(3 c_3-1\right)}{9 \left(c_3+1\right)} \,,\quad  \textup{and} \quad \kappa^2 > 0  \,, \\
 &   \frac16 \le c_3 < \frac13 \,, \quad  w > 0  \,,\quad \textup{and}\quad  w \ne -\frac{2 \left(3 c_3-1\right)}{9 \left(c_3+1\right)} \,,\quad  \textup{and} \quad \kappa^2 > 0 \,, \\
 &  c_3 \ge \frac13 \,, \quad    w > 0 \,,\quad  \textup{and} \quad \kappa^2 > 0  \,,
\end{aligned}
\right. \label{cond232}
\ee
In the case (2.4) the stabilities of the Einstein static universe~(\ref{sol24}) require conditions
\be
 \left\{
\begin{aligned}
 & \frac{1}{15} < c_3 \le \frac{1}{3} \,, \quad  -\frac{9 c_3^2 -9 c_3 +4}{12} \le c_4 < \frac{27 c_3^2 +38 c_3 -21}{64} \,, \\
 & \frac{1}{3} < c_3 \le \frac{13}{27} \,, \quad  -\frac{9 c_3^2 -9 c_3 +4}{12} \le c_4 < \frac{27 c_3^2 +38 c_3 -21}{64}  \,,\quad c_4 \ne -\frac{c_3}{4}  \,, \\
 &    \frac{13}{27} < c_3 < \frac{2}{3} \,, \quad -\frac{9 c_3^2 -9 c_3 +4}{12} \le c_4 < \frac{27 c_3^2 +38 c_3 -21}{64}   \,,\quad c_4 \ne -\frac{c_3}{4}  \,, \quad  c_4 \ne \frac{5-9 c_3}{12} \,, \\
 &   \frac{2}{3} \le c_3 < \frac{5}{6} \,, \quad -\frac{ c_3 }{4} < c_4 < \frac{27 c_3^2 +38 c_3 -21}{64}   \,,\quad \textup{and}\quad  c_4 \ne \frac{5-9 c_3}{12}  \,, \\
 &  c_3 = \frac{5}{6} \,, \quad -\frac{5}{24} < c_4 <\frac{353}{768} \,,   \\
 &  c_3 > \frac{5}{6} \,, \quad -\frac{ c_3 }{4} < c_4 < \frac{27 c_3^2 +38 c_3 -21}{64}   \,,
\end{aligned}
\right. \label{cond241}
\ee
and
\be
w= -1 \,, \quad 0 \le \kappa^2 <  m^2 a_m^2 \sqrt{ 9 c_3^2 -9 c_3  +12 c_4 +4} \,. \label{cond242}
\ee
For conditions~(\ref{cond232}) in case (2.3), the solution~(\ref{sol23}) can be stable against inhomogeneous scalar perturbations, rather homogeneous perturbation. For conditions~(\ref{cond21})-(\ref{cond231}), (\ref{cond241}) and (\ref{cond242}) in cases (2.1)-(2.4), the Einstein static flat universes can be stable only against homogeneous scalar perturbation $(\kappa^2=0)$. Strictly, these solutions might not be stable against inhomogeneous scalar perturbations because $a_m$ and $m$ cannot go to infinity for $\kappa \rightarrow \infty$. We could only say the Einstein static flat universes can be stable against both homogeneous and some modes of inhomogeneous scalar perturbations filled with a cosmological constant $(w=-1)$, quintessence $(-1 <w < -1/3)$ and phantom $(w<-1 )$, and suitable parameters $c_3,$ and $c_4$. However, the  Einstein static flat universe cannot be stable under homogeneous and complete inhomogeneous scalar perturbations.


\subsubsection{Case 3: $e_3 \ne 0$}\label{sec324}

\begin{figure}[ht]
\centerline{
\includegraphics[width=0.3\linewidth]{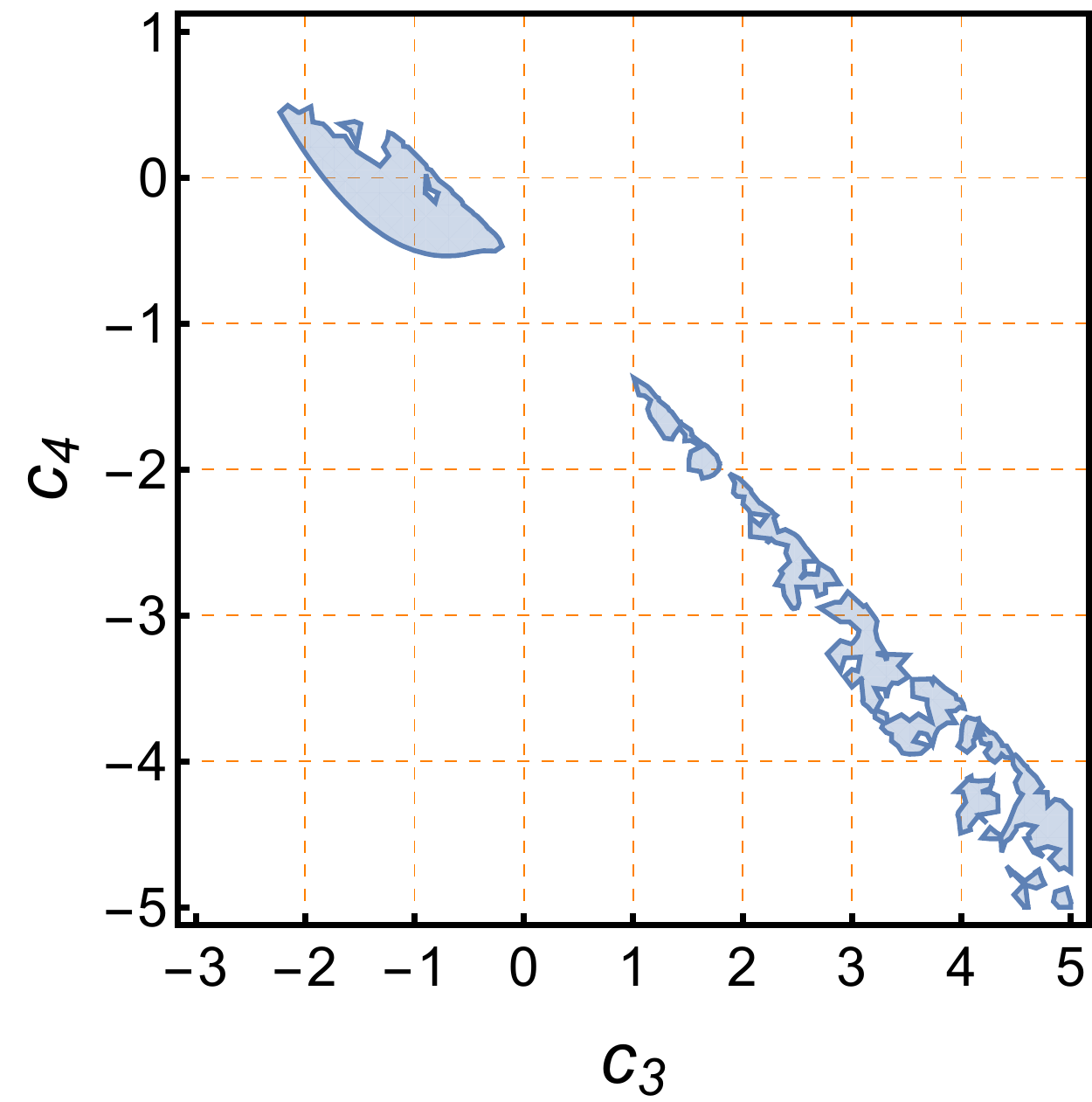}\ \ \ \ \ \includegraphics[width=0.3\linewidth]{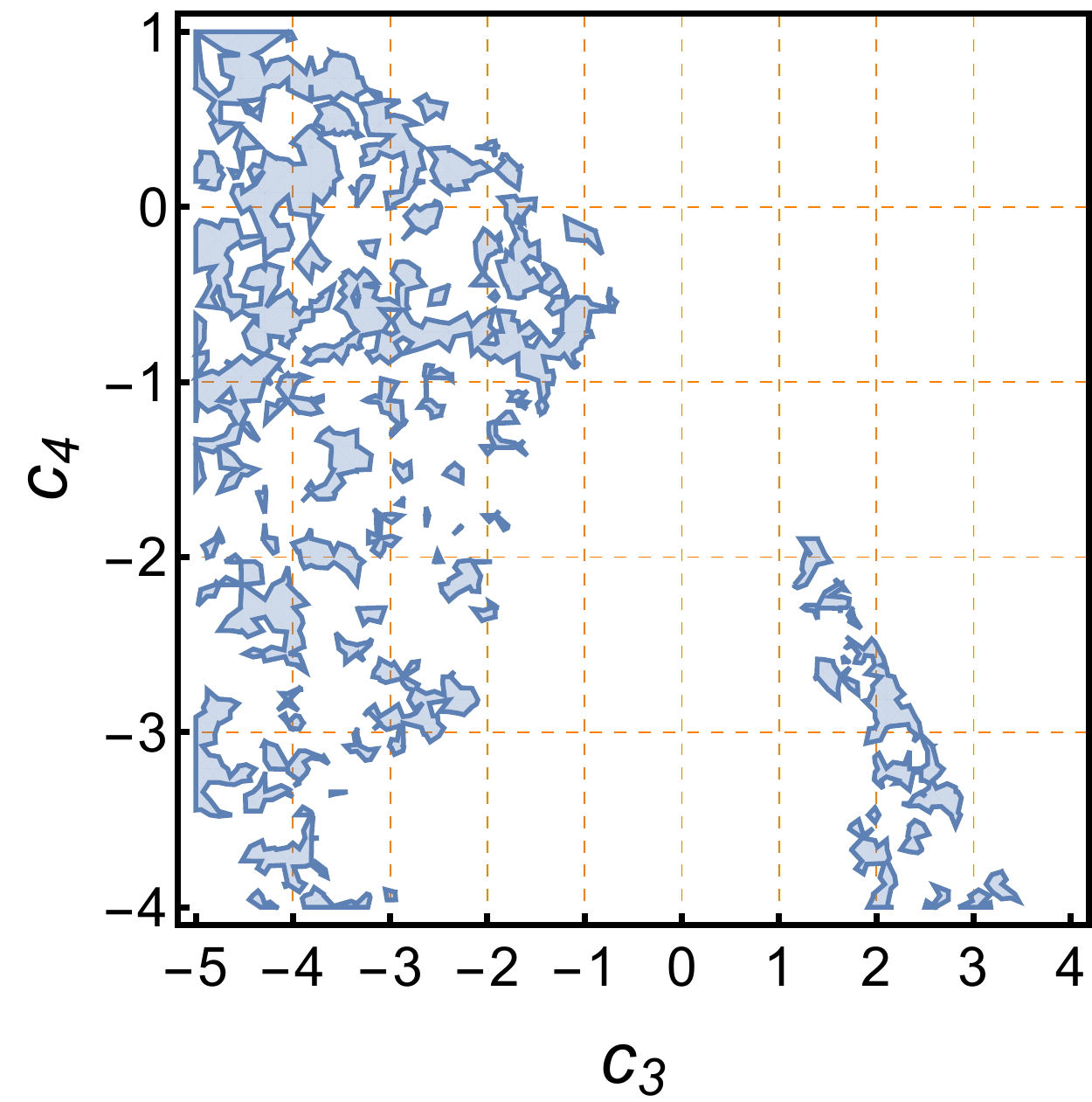}\ \ \ \ \ \includegraphics[width=0.3\linewidth]{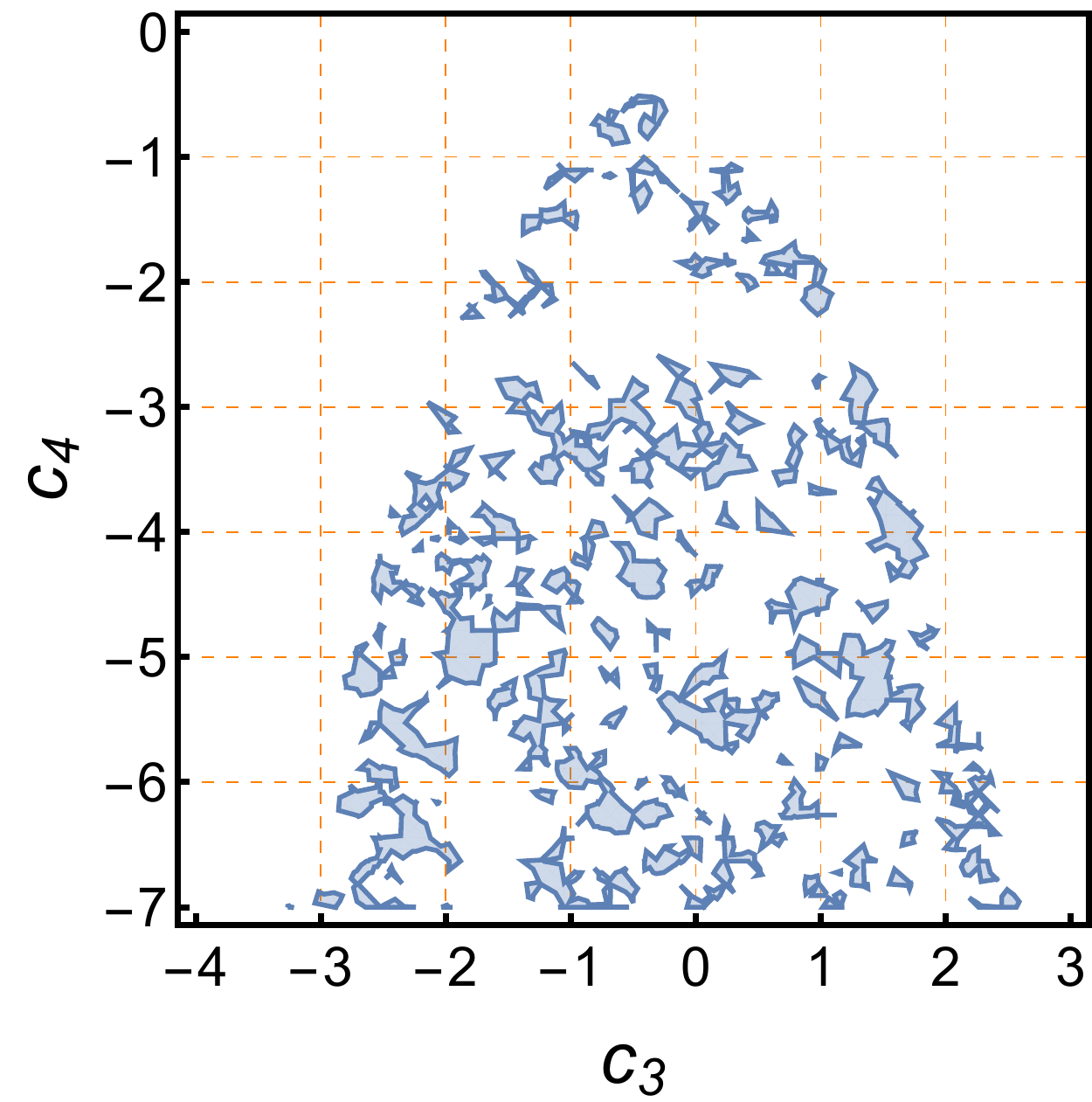}}
\caption{\small The parameters $(c_3, c_4)$ region of the stable Einstein static flat solutions $n_2 $ (left), $n_3$ (middle) and $n_4$ (right) against both homogeneous and inhomogeneous scalar perturbations $\kappa^2 \ge 0 $.  We set $m=1$ and $a_m =100$ for simplicity.}
\label{solfig2}
\end{figure}

In  this case, we also study  the parameters $(c_3, c_4, w)$ region  of the stabilities conditions numerically. However, in order to obtain the stable Einstein static universe against inhomogeneous scalar perturbations, we should consider all modes of the perturbations, i.e. $\kappa^2 = 1, 4, 9 \dots \infty $. It is not easy to analyze numerically.  So we study the stabilities in some special cases.  According to the stability condition~(\ref{Zcond}), we find that $\kappa^2$ does not impact the condition when $w = 0$. And the case $w=0$ represents the Universe is filled with ordinary matter, which is important and received with great interests. We find that the stable Einstein static flat universes filled with ordinary matters $w=0$ do exist. And we plot the parameters $(c_3, c_4)$ regions of the stable solutions in $w=0$ cases in Fig.~\ref{solfig2}
 for simplicity.
 For a concrete demonstration, we choose
 \be
 c_3 = -2 \,, \quad c_4 = \frac{1}{5} \,,\quad w=0 \,,\quad n= n_2 \approx 0.18 \,.
 \ee
 It is worth noting that, to our  knowledge, our construction is the first of the stable Einstein static universes with the flat spatial geometry,  in the presence of ordinary matter against both homogeneous and inhomogeneous scalar perturbations in modified gravities.


\section{Bouncing and cyclic universes}\label{sec4}

In Ref.~\cite{Feng:2018qnx}, the cosmological time crystal with two jumping points was constructed to realize bouncing universes, and cyclic universes in degenerate massive gravity together with Einstein -Riemann cubic gravities and some matters. These  cosmological time-crystal solutions are characterized by the discontinuity of $\dot a$ at the turning points.
In this section, we would like to turn off the higher-order derivative terms and construct the smooth bouncing and cyclic models in degenerate massive gravity. To be specific, we focus on the construction of classes of exact solutions of bouncing universes, and cyclic universes.  The total action ${\cal S}$ is given by Eq.~(\ref{action}). The gravitational part is still degenerate massive gravity. However, the action of matter ${\cal S}_m$ is given by
\begin{gather}
{\cal S}_m = \int d^4 x \sqrt{-g} L_m  \,, \quad \textup{with} \quad L_m =  -2 \Lambda_0 -  \sum_{i=1}^{3} \frac12 (\partial \varphi_i)^2 \,,  \\
\varphi_1 = 2 \alpha x \,,\quad \varphi_2 = 2 \alpha y \,,\quad \varphi_3 = 2 \alpha z \,,
\end{gather}
where $\Lambda_0$ is the bare cosmological constant. Note that we further added three axion fields $\varphi_i$ with a positive constant $\alpha$. These axions preserve the homogeneity and isotropicity of the background cosmological metric, but can have nontrivial perturbative effects~\cite{Geng:2014vza}. (These matter fields also can be used to construct cosmological time crystals in the presence of higher-order derivative gravity.) 
As we shall see presently, the axions are not essential but optional for constructing bouncing and cyclic models; we include them nevertheless for presenting a bigger theory. The effective Lagrangian $L$ for the FLRW metric~(\ref{frw}) is given by
\be
L = - 6 a \dot{a}^2 -V \,, \label{efflag} \\
\ee
with
\be
\begin{split}
V&= 2 \left(\Lambda_0 -6 m^2 (2 c_3+2 c_4+1)\right) a^3 +18 a_m m^2 (3 c_3+4 c_4+1) a^2 +6 (\alpha^2  - a_m^2 m^2 (6 c_3 \\
&\quad  +12 c_4+1)) a +6 a_m^3 m^2 (c_3+4 c_4) \,.
\end{split}
\ee
The corresponding Hamiltonian constraint is given by
\be
H= - 6 a \dot{a}^2 +V =0 \,. \label{effham}
\ee
which can be viewed as the effective equation of motion. And it can be rewritten as a differential equation,
\be
\dot{a}^2 +w_1 a^2 +w_2 a +w_3 +\frac{ w_4}{a} =0  \,, \label{bouncecycliceq}
\ee
with
\be
\begin{split}
w_1 &= -\frac{\Lambda_0}{3} +2( 1+2 c_3+2 c_4)m^2  \,,\quad w_2 = -3 a_m (1+3 c_3+4 c_4) m^2 \,,\\
w_3 &= -\alpha^2 + a_m^2 (1+6 c_3 +12 c_4) m^2 \,, \quad w_4 = - a_m^3 (c_3+ 4 c_4) m^2 \,.
\end{split}
\ee
The Eq.~\ref{bouncecycliceq} admits classes of exact solutions of bouncing universes, and cyclic universes satisfying when we restrict the parameters to satisfy $w_4 =0$, i.e. $c_4 = -c_3/4$. For $c_4\ne -c_3/4$, our numerical analysis indicates that the bouncing universes, and cyclic universes also exist, but exact solutions are not presentable. Instead, we shall present the linearized cyclic solution as a perturbation of Minkowski spacetime.

\subsection{Bouncing universe }\label{sec41}

We consider that the initial state taking a cosh-type ansatz for a bouncing model,
\be
a = A_1 + A_2 \cosh A_3 t \,,\label{bouncingmodel}
\ee
where $A_1, A_2$, and  $A_3$ are constants and obey the following reality conditions,
\be
 A_2 > 0 \,,\quad 0<  A_1 +A_2 <a_m \,, \quad A_3>0 \,.
\ee
For the FLRW metric~(\ref{frw}), the solutions are given by
\be
 A_1 = -\frac{w_2}{2 w_1} \,, \quad A_2 = -\frac{\sqrt{w_2^2 - 4 w_1 w_3}}{2 w_1} \,, \quad A_3 = \sqrt{-w_1}  \,.
\ee
The existence of the cosh-type bouncing solution requires that
\be
c_4 = -\frac{c_3}{4} \,, \quad  c_3 > -\frac13 +\frac{\alpha^2}{3 a_m^2 m^2} \,,\quad  \Lambda_0  > 3 (2 +3 c_3)  m^2   \,.
\ee
It is easy to see that the bouncing model can exist without the axions, i.e. $\alpha = 0$; however, the non-vanishing axions can modify the constraint on the parameters of massive gravity. On the other hand, the bare cosmological constant is necessary and positive in the construction of these bouncing solutions. As we shall see later, massive gravity itself provides repulsion at the bounce point and the theory can tolerate some attractive force without destroying the bounce.  However, additional repulsive force from positive cosmological constant must be included for the Universe not  to contract in the later time. It is worth noting that the bouncing model~(\ref{bouncingmodel}) can only describe the very early stage of the evolution of the Universe. The scale factor of the bouncing model will beyond the allowed max value $a_m$~\cite{DeFelice:2012mx} after a period of inflation. On the one hand, this problem should be solved in another stage of the evolution of the Universe. On the other hand, we can consider the Universe is oscillating. We study this case in the following subsection.

\subsection{Cyclic universe }\label{sec42}

We consider that the initial state taking a  sin-type ansatz for cyclic or oscillating universes:
\be
a = B_1 +B_2 \sin( B_3 t -\frac{\pi}{2}) \,, \label{cyclic}
\ee
where $B_1, B_2$, and  $B_3$ are constants and obey the following reality conditions,
\be
0<B_1\pm B_2<a_m \,, \quad B_1 > 0 \,, \quad B_2 \ne 0   \,, \quad B_3>0 \,.
\ee
For the FLRW metric~(\ref{frw}), the sin-type solutions are given by
\be
B_1 = -\frac{w_2}{2 w_1}  \,,\quad B_2 = \frac{ \sqrt{w_2^2- 4 w_1 w_3}}{2 w_1} \,, \quad B_3 = \sqrt{w_1} \,,
\ee
The existence of the sin-type cyclic solution requires that
\be
c_4 = -\frac{c_3}{4} \,, \quad c_3 > -\frac13  +\frac{\alpha^2}{3 a_m^2 m^2} \,,\quad \frac{3 m^2 ( 4\alpha^2 (2 +3 c_3) +a_m^2 m^2) }{4 ( \alpha^2 - (1 +3 c_3) a_m^2 m^2)} <\Lambda_0 < -\frac{3\alpha^2}{a_m^2} \,.
\ee
It can be seen that the cyclic model can also exist without  axions.  It is worth commenting that $\Lambda_0$, the bare cosmological constant, must be negative for these cyclic solutions, whilst it must be positive for the bounce solutions studied in the previous section. This is because the massive gravity by itself can provide sufficiently large repulsion to overcome the attraction from the negative bare cosmological constant for the universe to bounce; it requires a sufficient attractive force from the bare cosmological constant for the Universe to contract at a later time so that the Universe becomes cyclic.

\subsection{ Cyclic universe as linear perturbation}\label{sec43}

In the previous subsections, we consider $c_4 = -c_3/4$, for which exact solutions of bouncing universes, and cyclic universes could be obtained. We now consider the more general $c_4 \ne -c_3/4$ and we would like to construct cyclic universes whose $a$ can be viewed as a small perturbation from the Minkowski spacetime, and hence we can obtain the exact solution for the linearized metric.
In other words, we consider the Universe~(\ref{cyclic}) oscillating in a small range comparing with the lowest value of scale factor, i.e. $B_1 \gg | B_2 | $. The cyclic or oscillating ansatz can be rewritten as
\be
a(t) = C_1 + \epsilon  C_2 \,, \label{apert}
\ee
where the constant $C_1$ is the zeroth order solution, describing the Minkowski spacetime,  $C_2(t)$ is the first order solution, and $\epsilon$ is a small quantity. According to the Euler-Lagrangian equations,  the existence of  the zeroth order solution $C_1$ requires
\be
\frac{\partial V}{\partial a}\Big |_{a=C_1} =0 \,,
\ee
 and we have
\be
c_4 = \frac{\alpha^2 + C_1^2 \left(\Lambda_0 -6 \left(2 c_3+1\right) m^2\right)+6 C_1 \left(3 c_3+1\right) m^2 a_m-\left(6 c_3+1\right) m^2 a_m^2}{12 m^2 \left(C_1 -a_m\right)^2} \,. \label{c4cond}
\ee
We substitute Eqs~(\ref{apert}) and (\ref{c4cond}) into effective Lagrangian~(\ref{efflag}) and Hamiltonian~(\ref{effham}), and then perform a series expansion  of the effective Lagrangian and Hamiltonian to the second order. We find
\begin{align}
\begin{split}
L &= -V_0 + \big( -6 C_1 \dot{C}_2^{ 2} -\frac{6 C_2^2}{a_m -C_1} \big(   a_m^2 m^2 (2+3 c_3) +a_m  C_1 (\Lambda_0 -3 m^2 ( 1+c_3 )) \\
&\quad+ \alpha^2 \big) \big) \epsilon^2  + \mathcal{O} (\epsilon^3) \,,
\end{split} \\
\begin{split}
H &=  V_0 + \big( -6 C_1 \dot{C}_2^{ 2} +\frac{6 C_2^2}{a_m -C_1} \big(   a_m^2 m^2 (2+3 c_3) +a_m  C_1 (\Lambda_0 -3 m^2 ( 1+c_3 )) \\
&\quad+ \alpha^2 \big) \big) \epsilon^2  + \mathcal{O} (\epsilon^3) \,,
\end{split}
\end{align}
where $V_0$ is constant and given by
\be
V_0 = 2  \alpha^2 (a_m +2 C_1) -2 a_m ( a_m^2 m^2 ( 1+3 c_3) -2 a_m m^2 C_1 (2+3 c_3) -C_1^2 ( \Lambda_0 -3 (1+c_3) m^2) )  \,.
\ee
The vanishing of $V_0$ implies ghost instabilities, so we set it equal to a second order small quantity,
\be
V_0 = \epsilon^2 \lambda .
\ee
Note that $\lambda$ here can be of any sign and any finite constant, as long as $\epsilon$ perturbation is sufficiently small. According to the above equation, we have
\be
c_3 = \frac{2 \alpha^2 (2 C_1 +a_m) + 2 a_m \left(C_1^2 (\Lambda_0 -3 m^2 ) +4 C_1 a_m m^2 -a_m^2 m^2 \right) +\lambda \epsilon ^2}{6 m^2 \left(C_1 -a_m\right)^2 a_m} \,. \label{c3cond}
\ee
Substituting the above equation into the second order effective Lagrangian and Hamiltonian, we have
\bea
L_2 &=& -6 C_1 \dot{C}_2^2 -\frac{6 \left( \alpha^2 (2 a_m +C_1) + a_m^2 \left(C_1 (\Lambda_0 -m^2 )+m^2 a_m\right) \right) C_2^2}{\left(a_m - C_1 \right)^2} +\lambda \,, \\
H_2 &=& -6 C_1 \dot{C}_2^2 +\frac{6 \left( \alpha^2 (2 a_m +C_1) + a_m^2 \left(C_1 (\Lambda_0 -m^2 )+m^2 a_m\right) \right) C_2^2}{\left(a_m - C_1 \right)^2} -\lambda \,.
\eea
For $H_2 =0$, we can rewrite
\be
\dot{C}_2^2 +\hat{w}_1 {C}_2^2 -\hat{w}_2 =0 \,,\label{ham2eq}
\ee
with
\be
\hat{w}_1 = -\frac{\alpha^2 (2 a_m +C_1) + a_m^2 (C_1 (\Lambda_0 -m^2 )+m^2 a_m )}{C_1 (a_m -C_1 )^2} \,,\quad \hat{w}_2 = -\frac{\lambda}{6 C_1} \,.
\ee
Considering that we have the $\sin$-type oscillating ansatz, the solution is given by
\be
C_2 = \sqrt{\frac{\hat{w}_2}{\hat{w}_1}}  \sin \left(\sqrt{\hat{w}_1}\ t -\frac{\pi}{2}\right)\,,
\ee
which satisfies Eq.~(\ref{ham2eq}). The existence of the solution requires the following conditions,
\be
\hat{w}_1 >0 \,,\quad \hat{w}_2 >0  \,, \quad 0< C_1 \pm \epsilon \ C_2 <a_m  \,.
\ee
Finally, we have
\begin{align}
\begin{split}
& 0< C_1\le \frac{a_m}{2}  \,,\quad \Lambda_0<-\frac{(a_m -C_1) m^2}{ C_1 }\,,\quad 0\le \alpha < a_m \sqrt{\frac{C_1 m^2 -a_m m^2 -C_1 \Lambda_0}{2 a_m +C_1} } \,, \\
& \frac{6 C_1^2 ( \alpha^2 (2 a_m +C_1) +a_m^2 (C_1 (\Lambda_0 -m^2) +a_m m^2 )}{(a_m -C_1)^2 \epsilon ^2}< \lambda<0 \,,
\end{split} \\
\begin{split}
&\frac{a_m}{2} < C_1 < a_m \,,\quad \Lambda_0<-\frac{(a_m -C_1) m^2}{ C_1 }\,,\quad 0\le \alpha < a_m \sqrt{\frac{C_1 m^2 -a_m m^2 -C_1 \Lambda_0}{2 a_m +C_1} } \,, \\
&  \frac{6 ( \alpha^2 (2 a_m +C_1) +a_m^2 (C_1 (\Lambda_0 -m^2) +a_m m^2 )}{ \epsilon ^2}< \lambda<0    \,.
\end{split}
\end{align}
 $c_3$ and $c_4 $ are given by Eqs.~(\ref{c4cond}) and (\ref{c3cond}).   Similar to the previous case, the axions can be turned off, but the bare cosmological constant is necessary to realize the Universe oscillating in a small range.

\section{Conclusions and discussions}\label{sec5}

In this paper, we investigated massive gravity with degenerated reference metrics, focusing on the feasibility of some alternative inflationary models such as the emergent universe scenario, bouncing universes, and cyclic universes. 

We first studied the feasibility of the emergent universe scenario. We  constructed the Einstein static flat universe in degenerate massive gravity filled with perfect fluids. We then derived the linearized equations of motion in this background and studied the stabilities against both the homogeneous and inhomogeneous scalar perturbations. We found that there could exist stable such a universe filled with ordinary matter ($w=0$). Our construction is the first of the stable Einstein static universe with the flat spatial geometry,  in the presence of ordinary matter against both homogeneous and inhomogeneous scalar perturbations in modified gravities. The results show that the Einstein static flat universe can safely enter an inflationary epoch.  Our conclusion is significant since the universe with flat geometry appears to be favored by  latest astronomical observations~\cite{Bennett:2012zja, Aghanim:2018eyx}.

We also constructed classes of exact solutions of bouncing universes, and cyclic flat universes in degenerate massive gravity by including a bare cosmological constant and three free axion fields. It turns out that the cosmological constant is necessary but the axions are optional in the construction. For appropriate parameters, we found that cyclic universes could also emerge as some linear perturbations of the flat Minkowski spacetime. In our solutions, for the bounce universes, the bare cosmological constant must be positive whilst it must be negative for the cyclic universes.  In the latter case, the attractive force from the negative bare cosmological constant is necessary to overcome the repulsion from massive gravity to provide a contracting point so that the Universe becomes cyclic.
Our results demonstrate that bouncing universes, and cyclic universes can emerge in massive gravity coupled to a bare cosmological constant. The simplicity of the theory and the existence of such simple exact solutions open a new avenue to study alternative inflationary cosmology.

Our initial investigation of alternative cosmological models in degenerated massive gravity showed a new possibility of addressing cosmological problems. However, many works remain. All perturbations, including vector and tensor perturbations, should be analyzed when we study the stabilities of the Einstein static universe. Furthermore stabilities of bouncing and cyclic solutions should be also investigated. We leave these to future works.

\section*{Acknowledgement }

We are grateful to the anonymous referee for valuable comments. S.L.L. is grateful to Hua-Kai Deng, Xing-Hui Feng and Hyat Huang for useful discussions.  SLL and HW are  supported in part by NSFC grants No.~11575022 and No.~11175016, HL is supported in part by NSFC grants No.~11875200 and No.~11475024, PXW and HWY  are supported in part by NSFC grants No.~11435006, No.~11690034 and No.~11775077.


\end{document}